\documentclass[12pt]{amsart}
\usepackage{amscd}
\usepackage{amsfonts}
\usepackage{epsf}
\usepackage{epic}
\usepackage{eepic}
\usepackage{latexsym}

\begin{document}

\title[Quantum State Density in M-Theory]
{Quantum State Density and Critical Temperature in M-Theory}
 
\author{M.C.B. Abdalla}
\address{Instituto de F\'isica Te\'orica, Universidade Estadual Paulista,
Rua Pamplona 145, 01405-900 - S\~ao Paulo, SP, Brazil\,\, 
{\em E-mail address:} {\rm mabdalla@ift.unesp.br}}
\author{A.A. Bytsenko}
\address{Departamento de F\'isica, Universidade Estadual de Londrina,
Caixa Postal 6001, Londrina-Parana, Brazil; on leave from Sant-Petersburg
State Technical University, Russia \,\, {\em E-mail address:} 
{\rm abyts@uel.br}}
\author{B.M. Pimentel}
\address{Instituto de F\'isica Te\'orica, Universidade Estadual Paulista,
Rua Pamplona 145, 01405-900 - S\~ao Paulo, SP, Brazil\,\, 
{\em E-mail address:} {\rm pimentel@ift.unesp.br}}

\date{February, 2001}

\thanks{We would like to thank Prof. J.G. Russo for very constructive
discussion. A.A. Bytsenko gratefully acknowledges FAPESP's grant (Brazil), 
and the Russian Foundation for Basic Research, grant No. 01-02-17157. 
B.M. Pimentel thanks to CNPq for partial support.}

\maketitle

\begin{abstract}

We discuss  the asymptotic properties of quantum sta\-tes density for
fundamental $p-$ branes which can yield a microscopic interpretation
of the thermodynamic quantities in M-theory. The matching of the BPS 
part of spectrum for superstring and supermembrane gives the possibility 
of getting membrane's results via string calculations. In the weak coupling 
limit of M-theory the critical behavior coincides with the first order phase 
transition in standard string theory at temperature less then the Hagedorn's 
temperature $T_H$. The critical temperature at large coupling constant is 
computed 
by considering M-theory on manifold with topology 
${\mathbb R}^9\otimes{\mathbb T}^2$. 
Alternatively we argue that any finite temperature can be introduced in the 
framework of 
membrane thermodynamics.

\end{abstract}

\section{Introduction}

Recently deep connections between fundamental (super) membrane and (super) 
string 
theory have been found. In particular, it has been shown that the BPS 
spectrum of 
states for type IIB string on a circle is in correspondence with the BPS 
spectrum 
of fundamental compactified supermembrane \cite{schw95-360-13,russ96u-47}.
Membrane thermodynamics can indicate non trivial information about microscopic
degrees of freedom and the behavior of quantum systems at high temperature.  

Finite temperature M-theory defined on a manifold with topology 
${\mathbb R}^9\otimes{\mathbb T}^2$, at weak and strong string coupling 
constant 
regime, has been considered recently in Ref. \cite{russo}. In the small 
circle radius 
limit (radius of compactification) M-theory must recover the  string 
thermodynamics. 
In type IIB superstring theory, which is associated with M-theory in the 
weak coupling 
limit, the critical temperature coincides with the Hagedorn temperature 
\cite{russo}. 
In fact, in string theory there is a first order phase transition at 
temperature less 
than $T_H$ with  a large latent heat leading to a gravitational instability 
\cite{atick}. 
Note that the critical behavior in type IIA superstring theory, at large 
coupling limit,
is not well understood. 

The purpose of the present paper is to consider again the above mentioned 
problems, 
comparing strong and 
weak coupling regims by considering M-theory on 
${\mathbb R}^9\otimes{\mathbb T}^2$, 
where one of the sides of the torus ${\mathbb T}^2$ is the Euclidean time direction 
(fermions obey antiperiodic boundary conditions). We turned into the problem of asymptotic 
density of quantum states for fundamental $p$-branes already initiated in 
Refs. \cite{byts93,byts94,eliz94,byts96}. 

In Section 2 we consider asymptotic expansions of generating functions and thermodynamic 
quantities related to fundamental $p-$ branes. The light-cone Hamiltonian formalism for 
membranes wrapped on a torus is summarized in Section 3. The weak coupling limit of M-theory
is considered in Section 4. We calculate the one-loop free energy associated to strings (and 
$p-$ branes). The limit of strong coupling constant is analized in Section 5. In thermodynamics 
of M-theory on ${\mathbb R}^9\otimes {\mathbb T}^2$ we find the critical temperature, which  
coincides with Hagedorn temperature also obtained in Ref. \cite{russo}.
However we argue that a more interesting possibility allows for a finite temperature
to be introduced into the membrane theory. Finally we end up with some remarks.

\section{Asymptotics of quantum states in p-brane thermodynamics}

Let us consider multi-component versions of the classical generating
functions for partition functions, namely

$$
{\mathfrak G}_{\pm}(z)=\prod_{{\bf n}\in {\mathbb Z}^p/\{{\bf 0}\}}
\left[1\pm
\exp\left(-z\omega_{{\bf n}}({\bf a}, {\bf g})\right)\right]^{\pm q}
\mbox{,}
\eqno{(2.1)}
$$
where $z=y+2\pi ix$, $\Re z>0$,\, $q>0$ and $\omega_{{\bf n}}({\bf a}, 
{\bf g})$ is given by

$$
\omega_{\bf n}({\bf a}, {\bf g})=
\left(\sum_{j}a_j(n_j+{\rm g}_j)^2\right)^{1/2}
\mbox{,}
\eqno{(2.2)}
$$
${\rm g}_j$, and $a_j$ are some real numbers.

In the context of thermodynamics of fundamental $p-$ branes, classical generating 
functions ${\mathfrak G}_{\pm}(z)$ can be regarded as a partition function where 
$z\equiv\beta$ is the inverse temperature. 

In this paper we shall be working only with the temperature dependent part  of the 
free energy $F(\beta)$. In the following we write down the statistical free energy
 $F(\beta)$ (both for $p-$ branes and supersymmetric $p-$ branes), the internal 
 energy $E$ and the entropy $S$ 
 
$$
F_{p}(\beta)=-\frac{1}{\beta}{\rm log}\left[{\mathfrak G}_{-}(\beta)\right]
\mbox{,}
\eqno{(2.3)}
$$
$$
F_{sp}(\beta)=-\frac{1}{\beta}\mbox{log}\left[{\mathfrak G}_{+}(\beta)
{\mathfrak G}_{-}(\beta)\right]
\mbox{,}
\eqno{(2.4)}
$$
$$
E=\frac{\partial}{\partial\beta}\left[\beta F(\beta)\right],
\,\,\,\,\,\,\,\,\,\,
S=\beta^2\frac{\partial}{\partial\beta}F(\beta)
\mbox{.}
\eqno{(2.5)}
$$

However, in order to calculate the above quantities we need first to know 
the total number of quantum states which can be described by the quantities
$\Omega_{\pm}(n)$ defined by

$$
K_{\pm}(t)=\sum_{n=0}^{\infty}\Omega_{\pm}(n)t^n \equiv {\mathfrak G}_{\pm}(-\log t)
\mbox{,}
\eqno{(2.6)}
$$
where $t<1$, and $n$ is a total quantum number. The Laurent inversion formula
associated with the above definition has the form

$$
\Omega_{\pm}(n)=\frac{1}{2\pi \sqrt{-1}}\oint K_{\pm}(t)t^{-n-1}dt
\mbox{,}
\eqno{(2.7)}
$$
where the contour integral is taken on a small circle about the origin.

We shall use the results of Meinardus \cite{mein54-59-338,mein54-61-289,
andr76b} that can be easily generalized to the vector-valued functions of the
(2.1) type (for more detail see Ref. \cite{byts96}). 
The $p$-dimensional Epstein zeta function 
$Z_p\left|_{\bf h}^{\bf g}\right|(z,\varphi)$ associated with the quadratic 
form $\varphi [{\bf a}({\bf n}+{\bf g})]=(\omega_{\bf n}({\bf a}, {\bf g}))^2$ for 
$\Re\,z>p$ is given by the formula

$$
Z_p\left| \begin{array}{ll}
{\rm g}_1\,...\,{\rm g}_p \\
h_1\,...\,h_p\\
\end{array} \right|(z,\varphi)=\sum_{{\bf n}\in {\bf Z}^p}{}'
\left(\varphi[{\bf a}({\bf n}+{\bf g})]\right)^{-z/2}
$$
$$
\times\exp\left[2\pi \sqrt{-1}({\bf n},{\bf h})\right]
\mbox{,}
\eqno{(2.8)}
$$
where $({\bf n},{\bf h})=\sum_{i=1}^p
n_ih_i$,\, $h_i$ are real numbers and the prime on $\sum {'}$
means to omit the term ${\bf n} =-{\rm {\bf g}}$ if all the ${\rm g}_j$ are 
integers. For $\Re z<p$,\, $Z_p\left|_{\bf h}^{\bf g}\right|(z,\varphi)$ is
understood to be the analytic continuation of the right hand side of the
Eq. (2.6). The functional equation for 
$Z_p\left|_{\bf h}^{\bf g}\right|(z,\varphi)$ reads

$$
Z_p\left| \begin{array}{ll}
{\bf g}\\
{\bf h}\\
\end{array} \right|(z,\varphi)=
({\rm det}\,{\bf a})^{-1/2}\pi^{\frac{1}{2}(2z-p)}\frac{\Gamma(\frac{p-z}{2})}
{\Gamma(\frac{z}{2})}\exp[-2\pi\sqrt{-1}({\bf g},{\bf h})]
$$
$$
\times
Z_p\left| \begin{array}{ll}
{\bf\,\,\,\, h}\\
-{\bf g}\\
\end{array} \right|(p-z,\varphi^*)
\mbox{,}
\eqno{(2.9)}
$$
and $\varphi^*[{\bf a}({\bf n}+ {\bf g})]=\sum_j a_j^{-1}(n_j+{\rm g}_j)^2$.
Formula (2.9) gives the analytic continuation of the zeta function. 
Note that $Z_p\left|_{\bf h}^{\bf g}\right|(z,\varphi)$ is an entire 
function in the complex $z-$ plane except for the
case when all the $h_i$ are integers. In this case 
$Z_p\left|_{\bf h}^{\bf g}\right|(z,\varphi)$ has a simple pole
at $z=p$ with residue

$$
A(p)=\frac{2\pi^{p/2}}{({\rm det}\,{\bf a})^{1/2}\Gamma(p/2)}
\mbox{,}
\eqno{(2.10)}
$$
which does not depend on the winding numbers ${\rm g}_i$. Furthermore one has
$Z_p\left|_{\bf h}^{\bf g}\right|(0,\varphi)=-1$.

\medskip
\medskip

\par \noindent

{\bf Proposition 2.1.}\,\,\,\,\, {\em In the half-plane $\Re z>0$ there exists 
an asymptotic
expansion for ${\mathfrak G}_{\pm}(z)$ uniformly in $x$ as $y\rightarrow 0$,
provided $|{\rm arg}\,z|\leq \pi/4$ and $|x|\leq 1/2$ and given by

$$
{\mathfrak G}_{+}(z)=\exp\left\{q\left[A(p)\Gamma(p)\zeta_R(1+p)z^{-p}-
Z_p\left|_{{\bf 0}}^{{\bf {\rm g}}}\right|(0,\varphi){\rm log}2 +
{\mathcal O}\left(y^{c_{+}}\right)\right]\right\}
\mbox{,}
\eqno{(2.11)}
$$

$$
{\mathfrak G}_{-}(z)=\exp\left\{q\left[A(p)\Gamma(p)\zeta_{+}(1+p)z^{-p}-
Z_p\left|_{{\bf 0}}^{{\bf g}}\right|(0,\varphi){\rm log}z
\right. \right.
$$
$$
\left. \left.
+
(d/dz)Z_p\left|_{{\bf 0}}^{{\bf g}}\right|(z,\varphi)|_{(z=0)}
+{\mathcal O}\left(y^{c_{-}}\right)\right]\right\}
\mbox{,}
\eqno{(2.12)}
$$
where $0<c_{+},c_{-}<1$ and $\zeta_{-}(s)\equiv \zeta_R(s)$ is the Riemann 
zeta function, $\zeta_{+}(s)=(1-2^{1-s})\zeta_R(s)$.}

\medskip
\medskip
\medskip

By means of the asymptotic expansion of $K_{\pm}(t)$ for $t\rightarrow 1$, 
which is equivalent to the expansion of ${\mathfrak G}_{\pm}(z)$ for small 
$z$,  
and using formulae (2.11) and (2.12) one arrives at a complete asymptotic 
limit of 
$\Omega_{\pm}(n)$:

\medskip
\medskip

\par \noindent

{\bf Theorem 2.1} (Meinardus; also see Ref. \cite{byts96}).\,\,\,\,\, 
{\em For $n \rightarrow \infty$ one has

$$
\Omega_{\pm}(n)=C_{\pm}(p)n^{\left(2q Z_p\left|_{{\bf 0}}^{{\bf g}}\right|
(0,\varphi)-p-2\right)/(2(1+p))}
$$
$$
\times\exp\left\{\frac{1+p}{p}[q A(p)\Gamma(1+p)\zeta_{\pm}(1+p)]^
{1/(1+p)}n^{p/(1+p)}\right\}[1+{\mathcal O}(n^{-\kappa_{\pm}})]
\mbox{,}
\eqno{(2.13)}
$$
$$
C_{\pm}(p)=[q A(p)\Gamma(1+p)\zeta_{\pm}(1+p)]^{(1-2q 
Z_p\left|_{{\bf 0}}^{{\bf g}}\right|(0,\varphi))/(2p+2)}
$$
$$
\times
\frac{\exp\left[q (d/dz)
Z_p\left|_{{\bf 0}}^{{\bf g}}\right|(z,\varphi)|_{(z=0)}\right]}
{[2\pi(1+p)]^{1/2}}
\mbox{,}
\eqno{(2.14)}
$$
$$
\kappa_{\pm}=\frac{p}{1+p}\min \left(\frac{C_{\pm}}{p}-\frac{\delta}{4},
\frac{1}{2}-\delta\right)
\mbox{,}
\eqno{(2.15)}
$$
and\,\, $0<\delta<\frac{2}{3}$.}

\medskip
\medskip
\medskip

Using  Eqs. (2.13) of the Theorem 2.1 and assuming linear Regge 
trajectories, i.e.
the mass formula $M^2=n$ for the number of brane states of mass $M$ to $M+dM$, 
one can obtain the asymptotic density for $p-$ brane states as well as for super $p-$ branes 
and they read 

$$
\varrho(M)dM\simeq 2C_{\pm}(p)M^{\left(1+2p-
2qZ_p\left|_{{\bf 0}}^{{\bf g}}\right|(0,\varphi)\right)/(1+p)}
\exp\left[b_{\pm}(p)M^{\frac{2p}{1+p}}\right]
\mbox{,}
\eqno{(2.16)}
$$
$$
b_{\pm}(p)\equiv\left(1+\frac{1}{p}\right)\left[q A(p)\Gamma(1+p)
\zeta_{R}(1+p)\right]^{\frac{1}{1+p}}
\mbox{.}
\eqno{(2.17)}
$$

$$
b_{sp}(p)\equiv\left(1+\frac{1}{p}\right)\left[q A(p)\Gamma(1+p)
(\zeta_{+}(1+p)+\zeta_{-}(1+p))\right]^{\frac{1}{1+p}}
\mbox{.}\eqno{(2.18)}
$$
In the supersymmetric case we had to  deal with product of generating
functions ${\mathfrak G}_{+}(\beta)\times{\mathfrak G}_{-}(\beta)$.

This result has a universal character for all $p-$ branes.
With the help of Theorem 2.1. we can compute 
the complete $p-$ brane state density (2.16), including the 
prefactors ${C}_{\pm}(p)$ and the factors $b_{\pm}(p)$, depending on 
the dimension of the embedding space. 

There are branes which do no wind around the torus. Those cannot be
treated in the semiclassical approximation. The free energy of the zero 
winding sector, for infinitely small Planck length, can be associated with
the energy of supergravity theory. Our goal, however, is the calculation 
of the temperature dependent state density of branes.

An attempt to compare the asymptotic states density of branes and the density 
of states of neutral black holes has been made in Ref. \cite{byts93, byts94}. 
The prefactor for the degeneracy of black hole states at mass level represents 
general quantum field corrections to the state density. The asymptotic behaviour 
of classical entropy of near-extremal black branes coincides  with the asymptotic 
degeneracy of some weakly interacting fundamental $p$-brane excitation modes.

The asymptotic density of states is consistent with the entropy of near-extremal 
$p$-branes (indeed $S\sim M^{\frac{2p}{p+1}}$). In the limit $\beta\rightarrow 0$ 
$(T\rightarrow\infty)$ the entropy of fundamental
objects may be identified with ${\rm log}(\Omega_{\pm}(n))$, while the 
internal energy is related to $n$. Therefore, from Eqs. (2.13), (2.14) and 
(2.3) - (2.5) one has

$$
F_{p}(T)\simeq -qA(p)\Gamma(p)\zeta_{R}(1+p)T^{p+1}
\mbox{,}
\eqno{(2.19)}
$$
$$
F_{sp}(T)\simeq -qA(p)\Gamma(p)\left[\zeta_{+}(1+p)+\zeta_{-}(1+p)
\right]T^{p+1}
\mbox{,}
\eqno{(2.20)}
$$
$$
E_{p}\simeq pq A(p)\Gamma(p)\zeta_{R}(1+p)T^{p+1}
\mbox{,}
\eqno{(2.21)}
$$
$$
E_{sp}\simeq pq A(p)\Gamma(p)\left[\zeta_{+}(1+p)+\zeta_{-}(1+p)\right]
T^{p+1}
\mbox{,}
\eqno{(2.22)}
$$
$$
S_{p}\simeq (1+p)q A(p)\Gamma(p)\zeta_{R}(1+p)T^p
\mbox{,}
\eqno{(2.23)}
$$
$$
S_{sp}\simeq (1+p)q A(p)\Gamma(p)\left[\zeta_{+}(1+p)+\zeta_{-}(1+p)\right]
T^p
\mbox{.}
\eqno{(2.24)}
$$
Eliminating the quantity $T$ among equations above one gets

$$
S_{p}\simeq\frac{1+p}{p}\left[q pA(p)\Gamma(p)\zeta_{R}(1+p)\right]^
{\frac{1}{1+p}}
E_{p}^{\frac{p}{1+p}}
\mbox{,}
\eqno{(2.25)}
$$
$$
S_{sp}\simeq\frac{1+p}{p}\left[q pA(p)\Gamma(p)\left(\zeta_{+}(1+p)+
\zeta_{-}(1+p)
\right)\right]^{\frac{1}{1+p}}E_{sp}^{\frac{p}{1+p}}
\mbox{.}
\eqno{(2.26)}
$$
Thus the behavior of the entropy can be understood in terms of the degeneracy of some
interacting fundamental $p-$ brane excitation modes. Generally speaking the
$p-$ brane approach can yield a microscopic interpretation of the entropy.

\section{Toroidal membranes}

In this section we will consider the light-cone Hamiltonian formalism
for membranes wrapped on a torus in Minkowski space. Such a 
compactification of M-theory with (-,+) spin structure, having the topology
${\mathbb R}^{9}\otimes {\mathbb T}^2$, assumes that the dimensions $X^{11},\,
X^{10}$ are compactified on a torus with radii $R_{10},\, R_{11}$
and two spatial membrane directions wind around this torus.

The single-valued functions on the torus $X^{10}(\sigma,\rho),
X^{11}(\sigma.\rho)$, where $\sigma ,\rho \in [0,2\pi )$ are the membrane 
world-volume coordinates, have the form 

$$
X^{10}(\sigma ,\rho )=m_0R_{10}\sigma + \widetilde X^{10} (\sigma ,\rho )\ ,
\ \ \  \ 
X^{11} (\sigma ,\rho )=R_{11} \rho  + \widetilde X^{11} (\sigma ,\rho )
\mbox{.}
\eqno{(3.1)}
$$
The eleven bosonic coordinates are
$\{ X^0, X^i,X^{10},X^{11}\}$ and in addition the transverse coordinates 
$X^i(\sigma ,\rho )$, $i=1,2,...,8$ are all single-valued.
Transverse coordinates can be expanded in a complete basis 
of functions on the torus, namely

$$
X^i(\sigma,\rho )=\sqrt{\alpha '} \sum_{k,\ell} X^i_{(k,\ell)} e^{ik\sigma +
i\ell\rho }
\mbox{,}
\eqno{(3.2)}
$$
$$
P^i(\sigma,\rho )=\frac{1}{(2\pi )^2 \sqrt{\alpha '}} \sum_{k,\ell} 
P^i_{(k,\ell)} 
e^{ik\sigma +i\ell\rho }
\mbox{.}
\eqno{(3.3)}
$$
Here $\alpha'=\big( 4\pi ^2 R_{11} T_{2}\big)^{-1}$, while
$T_{2}$ is the membrane tension. 

The membrane Hamiltonian in light-cone formalism 
\cite{berg,wit,russo1,russo2,russo}
can be written as follows: $H=H_0+H_{\rm int}$. In particular, for bosonic
modes of membrane the explicit Hamiltonian is 

$$
\alpha'  H_0= 8\pi^4 \alpha 'T_{2}^2 R_{10} ^2 R_{11}^2 m^2+ \frac{1}{2} 
\sum _{\bf n} 
\big[ P_{\bf n}^i P^i_{-{\bf n}}
+\omega_{km}^2 X^i_{\bf n} X^i_{-{\bf n} }\big]
\mbox{,}
\eqno{(3.4)}
$$

$$
\alpha'  H_{\rm int}= \frac{1}{4{\rm g}^2_A}\sum ({\bf n}_1 
\times {\bf n}_2)({\bf n}_3\times {\bf n}_4)
X_{{\bf n}_1}^i  X_{{\bf n}_2 }^j  X_{{\bf n}_3}^i   X_{{\bf n}_4}^j
\mbox{.}
\eqno{(3.5)}  
$$
In Eqs. (3.4) and (3.5) ${\bf n}\equiv (k,\ell)$, 
${\bf n}\times\ {\bf n}'=k\ell'-\ell k'$,
${\rm g}^2_A \equiv R_{11}^2(\alpha')^{-1}=4\pi^2R_{11}^3T_{2}$, 
$\omega_{k\ell} =(k^2 + m^2 \ell^2 R_{10}^2R_{11} ^{-2})^{1/2}$,
and $(m,k,\ell,k',\ell')\in {\mathbb Z}$.
The interaction term (3.5) depends on the type IIA
string coupling ${\rm g}_A$. Mode operators, related to basic functions 
$X^{i}(\sigma,\rho),\,P^i(\sigma,\rho)$, are

$$
X^i_{(k,\ell)}=\frac{1}{\sqrt {-2} \omega_{(k,\ell)} }
\big[\alpha^i_{(k,\ell)}+\widetilde\alpha^i_{(-k,-\ell)}\big],\,\,\,
P^i_{(k,\ell)}=\frac{1}{\sqrt {2} }\big[\alpha^i_{(k,\ell)}-
\widetilde \alpha^i_{(-k,-\ell)}\big]
\mbox{,}
\eqno{(3.6)}
$$
$$
\big( X_{(k,\ell)}^i\big) ^\dagger =X_{(-k,-\ell)}^i\ ,\ \ \ \ 
\big( P_{(k,\ell)}^i\big) ^\dagger =P_{(-k,-\ell)}^i
\mbox{,}
\eqno{(3.7)}
$$
and $\omega_{(k,\ell)}\equiv{\rm sign}(k )\,\omega_{k\ell}$.
The canonical commutation relations read

$$
\big[ X^i_{(k,\ell)} , P^j_{(k',\ell')} \big]= \sqrt{-1}\delta_{k+k'}
\delta_{\ell+\ell'}\delta^{ij}
\mbox{,}
\eqno{(3.8)}
$$

$$
[\alpha _{ {(k,\ell)} }^i , \alpha^j_{(k',\ell')}]= \omega_{(k,\ell)} 
\delta _{k+k'}\delta _{\ell+\ell'}\delta^{ij}
\mbox{.}
\eqno{(3.9)}  
$$
We have similar relations for the $\widetilde \alpha _{ {(k,\ell)} } ^i$.

Finally the  mass  operator takes the form 

$$
M^2=2p^+p^--(p^{i})^2-p_{10}^2=2 (H_0+H_{\rm int})-(p^{i})^2-p_{10}^2
\mbox{.}
\eqno{(3.10)}
$$

The Hamiltonian of membrane is non linear. But there are two situations 
where one can simplify this Hamiltonian:

\medskip
({\bf i}) {\em The limit ${\rm g}_A\rightarrow 0$.}

({\bf ii}) {\em The other limit of large ${\rm g}_A$.} 

\medskip
We shall consider these two cases in the next two sections.

\section{Zero torus area limit of M-theory (${\rm g}_A\rightarrow 0$)}

The zero torus area limit of M-theory on ${\mathbb T}^2$ is related to the
asymptotic ${\rm g}_A\rightarrow 0$ at fixed $(R_{10}/R_{11})$.
Such a limit in M-theory leads to a ten-dimensional type IIB string theory.
More precisely, it has been shown \cite{schw95-360-13,russo2}
that quantum states of M-theory describe the $(p,q)$ strings bound states of 
type IIB superstring theory. We will consider in this section string theory
at finite temperature associated with the 
${\rm g}_A\rightarrow 0$ limit of M-theory.

\subsection{Critical temperature in type II string theory}
Let us consider string theory in Euclidean space 
(time coordinate $X^0$ is compactified on a circle of circumference $\beta$).
The presence of coordinates  compactified on circles 
gives rise to winding string states. The string single-valued function 
$X^0(\sigma,\tau )$ admits an expansion:

$$
X^0(\sigma ,\tau )= x^0+ 2\alpha' p^0 \tau +2 R_0w_0\sigma + 
\widetilde X(\sigma,\tau )
\mbox{,}
\eqno{(4.1)}
$$
where $p^0=\ell_0(R_0)^{-1}$, \, $\ell_0,\,m_0 \in {\mathbb Z}$.
The Hamiltonian and the level matching constraints becomes

$$
H=\alpha' p_i^2 +\frac{m_0^2R_0^2}{\alpha' }+\alpha' \frac{\ell_0^2}{R^2_0}
+2(N_L+N_R-a_L-a_R)=0
\mbox{,}
\eqno{(4.2)}
$$
$$
N_L-N_R=\ell_0m_0
\mbox{,}
\eqno{(4.3)}
$$
where $a_L, a_R$ are the normal ordering constants, which 
represent the vacuum energy of the 1+1 dimensional field theory. 
In the case of type II superstring the number operators in the  
$m_0=\pm 1$ sector read 

$$
N_L=\sum_{{n}=1}^\infty \big[\alpha_{-{n}}^i \alpha _{n}^i +(n
-\frac{1}{2}) S_{-n}^a S^a_n \big]\ ,
\ \ \ N_R=\sum_{n=1}^\infty \big[\widetilde \alpha_{-n}^i \widetilde 
\alpha _n^i +(n-\frac{1}{2})
\widetilde S_{-n}^a \widetilde S^a_n\big]\ ,
\eqno{(4.4)}
$$
where $i=1,...,8,\,\,\,a=1,...,8$.
The normal-ordering constants are  the same as in the NS sector of the  NSR 
formulation, i.e. $a_L=a_R= 1/2$. 

The critical temperature of string can be obtained as usual by determining 
the radius $R_0$ at which appear infrared instabilities. In the following 
we reproduce the critical Hagedorn temperature  using the general state density 
formulation (2.16) - (2.18), which is more suitable for generalizing to membrane theory.
Indeed, in the string case $p=1$ the inverse critical temperature becomes:

$$
\beta_{cr} = \frac{1}{2}b_{sp}(1)=
\left[12\zeta_R(2)A(1)\right]^{1/2} = 
\left[2\pi^2 A(1)\right]^{1/2}
\mbox{,}
\eqno{(4.5)}
$$
where the the factor 1/2 is related to the right- and left- moving modes 
($N_L=N_R$) of closed string. Taking into account that $A(1)= 2$ and 
restoring the $\alpha'$ dependence one finds the critical Hagedorn's temperature

$$
T_{H}=\frac{1}{2\pi\sqrt{\alpha'}}
\mbox{.}
\eqno{(4.6)}
$$
This result is well known. In general the Hagedorn's temperature depends on the 
normal-ordering constants $a_L,\,\,a_R$ and has the form 
$T_H^{-1}=2\pi\sqrt{2\alpha'(a_L+a_R)}$.

\subsection{The one-loop free energy of strings}

Let us consider the semiclassical free energy associated with fundamental 
compactified (super) $p-$ branes (which is known to be
divergent) embedded in flat $D$-dimensional manifolds with topologies
${\mathfrak M}={\mathbb S}^1\otimes{\mathbb T}^p\otimes{\mathbb R}^{D-p-1}$. 
For the simplest quantum field model the free energy associated with bosonic 
and fermionic degrees of freedom has the form (see for example Refs.
\cite{odintsov,eliz94,byts96})

$$
F^{(b,f)}(\beta)=
- \pi^p({\rm det}{\mathfrak A})^{1/2}\int_0^{\infty}ds(2s)^{-(D-p+2)/2}\Xi^{(b,f)}
(s,\beta)
$$
$$
\times \Theta\left[\begin{array}{r}
{\bf g}\\
{\bf 0}
\end{array}\right]({\bf 0}|\Omega)
\exp{\left(-\frac{sM_0^2}{2\pi}\right)}
\mbox{,}
\eqno{(4.7)}
$$
where
$$
\Xi^{(b)}(s,\beta)=\theta_3\left(0\Big\vert\frac{\sqrt{-1}\beta^2}{2s}\right)-1
\mbox{,}\,\,\,
\,\,\,\Xi^{(f)}(s,\beta)=1-\theta_4\left(0\Big\vert\frac{\sqrt{-1}\beta^2}{2s}
\right)
\mbox{,}
\eqno{(4.8)}
$$
and $\theta_3(\nu|\tau)$ and $\theta_4(\nu|\tau)=
\theta_3(\nu+\frac{1}{2}|\tau)$ are the Jacobi theta functions.
Here ${\mathfrak A}={\rm diag}(R_1^{-2},...,R_p^{-2})$ is a $p\times p$ matrix. The
global parameters $R_j$ characterizing the non-trivial topology of ${\mathfrak M}$
appear in the theory due to the fact that coordinates $x_j (j=1,...,p)$ obey
the conditions $0\leq x_j<2\pi R_j$. The number of topological configurations
of quantum fields is equal to the number of elements in group $H^1({\mathfrak M};
{\mathbb Z}_2)$, that is, the first cohomology group with coefficients in ${\mathbb Z}_2$. 
The multiplet ${\bf g}=({\rm g}_1,...,{\rm g}_p)$ defines the topological type
of field (i.e., the corresponding twist), and depends on the field type
chosen in ${\mathfrak M}$, ${\rm g}_j=0$ or $1/2$. In our case 
$H^1({\mathfrak M};{\mathbb Z}_2)={\mathbb Z}_2^p$ and so the number of 
topological configurations of real scalars (spinors) is $2^p$. 

We follow the notations and treatment of Ref.
\cite{mumf84} and introduce the theta function with characteristics
${\bf a}, {\bf b}$ for ${\bf a},{\bf b}\in{\mathbb Z}^p$,

$$
\Theta\left[\begin{array}{r}
{\bf a}\\
{\bf b}
\end{array}\right]({\bf z}|\Omega)=\sum_{{\bf n}\in {\bf Z}^p}\exp \left[
\pi\sqrt{-1}({\bf n}+{\bf a})\Omega({\bf n}+{\bf a})
\right.
$$
$$
\left.
+ 2\pi \sqrt{-1}({\bf n}+{\bf a})({\bf z}+
{\bf b})\right]
\mbox{.}
\eqno{(4.9)}
$$
In this connection $\Omega=(s\sqrt{-1}/2\pi^2)\mbox{diag}(R_1^2,...,R_p^2)$. 

The above mentioned method of the free energy calculation admits a subsequent development for
extended objects. We will assume that the free energy is equivalent to a sum of
the free energies of quantum fields which are present in the modes of a $p-$ brane.
The factor $\exp(-sM_0^2/2\pi)$ in Eq. (4.7) should be understood as
$\mbox{Tr}\exp(-sM^2/2\pi)$, where $M$ is the mass operator of the brane and the
trace is taken over an infinite set of Bose-Fermi oscillators $N^{(b)}_{{\bf n}},
N^{(f)}_{{\bf n}}$.

The one-loop free energy of fields contained in a (super) $p-$ brane
can be evaluated making use the Mellin-Barnes representation 
for the energy integral $F(\beta)=\int_{\Re\,s=c}ds{\mathfrak F}(s,D)
{\rm Tr}[M^2]^{(D-s)/2}$\,\, (see Ref.\cite{byts1}) and the mass operators

$$
{\rm Tr}\left[e^{-tM^2}\right] = \frac{1}{2\pi\sqrt{-1}}\int_{\Re\,s=s_0}
ds\Gamma(s){\rm Tr}[tM^2]^{-s}
\mbox{,}
\eqno{(4.10)}
$$

$$
{\rm Tr}[M^2]^{-s}=
\frac{1}{\Gamma(s)}
\int_0^\infty dtt^{s-1}
{\rm Tr}\left[e^{-tM^2}\right]
\mbox{,}
\eqno{(4.11)}
$$

$$
{\rm Tr}\left[e^{-tM_p^2}\right]={\mathfrak G}_{-}(t),
\,\,\,\,\,\,\,\,\,\,\,\,
{\rm Tr}\left[e^{-tM_{sp}^2}\right]={\mathfrak G}_{+}(t){\mathfrak G}_{-}(t)
\mbox{,}
\eqno{(4.12)}
$$

$$
{\mathfrak G}_{\pm}(t)=\prod_{{\bf n}\in {\mathbb Z}^p/\{{\bf 0}\}}
\left[1\pm
\exp\left(-t\omega_{{\bf n}}\right)\right]^{\pm(D-p-1)}
\mbox{.}
\eqno{(4.13)}
$$
One can use some substraction procedure for the divergent terms in
${\mathfrak G}_{-}(t), \,{\mathfrak G}_{+}(t)$ in order to procede regularization 
(see for detail Ref. \cite{byts1}). To simplify the calculation we put $a_j=R_j=1$ and the  
final result for the free energy \cite{actor,byts2} is:

$$
F_{p}(\beta)\simeq -Q(D,p)
\sum_{k=1}
^{\infty}\frac{\Gamma\left(pk+\frac{1-p}{2}\right)}{\Gamma(k)}
\zeta_{R}(2pk+1-p)
x^{1+p(2k-1)} 
\mbox{,}
\eqno{(4.14)}
$$

$$
F_{sp}(\beta)\simeq -Q(D,p)
\sum_{k=1}
^{\infty}\frac{\Gamma\left(pk+\frac{1-p}{2}\right)}{\Gamma(k)}
\zeta_{+}(2pk+1-p)
x^{1+p(2k-1)}
\mbox{,}
\eqno{(4.15)}
$$
where $Q(D,p)$ is an integer function and

$$
x=\beta^{-1} y(p) \equiv \beta^{-1}
\left[(D-p-1)2^{3p-2}\pi^{\frac{p-1}{2}}\Gamma\left(\frac{p+1}{2}\right)
\zeta_R(p+1)\right]^{\frac{1}{2p}}
\mbox{.}
\eqno{(4.16)}
$$
The first terms of the leading behavior of the series (4.14), (4.15) ($k=1$) 
have to 
coincide with formulae (2.19) and (2.20) at $q=D-p-1$. Therefore we have

$$
Q(D,p)=(D-p-1)A(p)\Gamma(p)[y(p)]^{-1-p}
\mbox{.}
\eqno{(4.17)}
$$
The asymptotic expansion of $\Gamma(s)$ for large value of $|s|$ has the form

$$
\Gamma(s) =s^{s-\frac{1}{2}}e^{-s}\sqrt{2\pi}
\left(1+{\mathcal O}(s^{-1})\right),
\,\,\,\,\,\,\,|{\rm arg}\,s|<\pi
\mbox{,}
\eqno{(4.18)}
$$
and for $p>1$ the power series (4.14) and (4.15) are divergent for any 
$x>0$.

In the string case ($p=1$) the corresponding series in Eqs. (4.14) and (4.15) can be
resummed into trigonometrical form using the identities

$$
\sum_{k=1}^{\infty}\zeta_R(2k)x^{2k}=\frac{1}{2}-\frac{1}{2}\pi x 
{\rm cot}(\pi x)
\mbox{,}
\eqno{(4.19)}
$$

$$
\sum_{k=1}^{\infty}\zeta_R(2k)\left(1-2^{-2k}\right)x^{2k}=
\frac{\pi x}{4}{\rm tan}\left(\frac{\pi x}{2}\right)
\mbox{.}
\eqno{(4.20)}
$$
The finite radius of Laurent series convergence $|x|<1$ corresponds to the
Hagedorn temperature in string thermodynamics (see for detail 
Ref. \cite{byts1}). Using trigonometric relations, formulae (4.14) and (4.15)
display a certain periodicity in temperature. The physical meaning of that
behaviour is still obscure.

\section{Large string coupling limit (${\rm g}_A\rightarrow \infty$)}

We now focus on the case ({\em ii}) mentioned in the end of section 3. by letting 
the constant ${\rm g}_A$ be large. In this limit $R_{10},\,R_{11}$ are large with fixed
$(R_{10}/R_{11})$ and the non linear interacting Hamiltonian is multiplied by the small 
constant ${\rm g}_A^{-2}$ so that it can be considered perturbatively. In the leading 
order of perturbative theory in ${\rm g}_A^{-2}$ the interaction term can be dropped. 
The solution to the membrane equations of motion has the form \cite{russo}

$$
X^i (\sigma, \rho, \tau )= x^i +\alpha'  p^i \tau + 
\sqrt{-\alpha'/2} \sum_{{\bf n} \neq (0,0)}
\omega_{\bf n}^{-1} 
$$
$$
\times \big[ \alpha _{\bf n}^i e^{ik\sigma +i\ell\rho } 
+ \widetilde \alpha _{\bf n} ^i e^{-ik\sigma -i\ell\rho }\big] 
\ e^{i w_{\bf n} \tau }
\mbox{.}
\eqno{(5.1)}
$$
The momentum components in the directions $X^{10}$ and $X^{11}$ are given by 
$p_{10}=(\ell_{10}/R_{10}),$ and $p_{11}=(\ell_{11}/R_{11})$,
where $\ell_{10},\,\ell_{11}\in {\mathbb Z}$. The nine-dimensional mass 
operator reads $M^2={\mathfrak H}$, where

$$
{\mathfrak H}= \frac{\ell^2_{10}}{R_{10}^2} + \frac{\ell^2_{11}}{R_{11}^2} + 
\frac{m_0^2 R_{10}^2}{ 
\alpha  ^{\prime 2}}\  + \frac{1}{\alpha' }H
\mbox{,}
\eqno{(5.2)}
$$
$$
H=  \sum _{k,\ell} \big( \alpha^i_{(-k,-m\ell} \alpha^i_{ (k,\ell)} 
+ \widetilde \alpha^i_{(-k,-\ell)} \widetilde \alpha^i_{(k,\ell)}\big)
\mbox{.}
\eqno{(5.3)}
$$
The level-matching conditions are \cite{duff,russo}

$$
N_\sigma^+ - N_\sigma^- = m_0 \ell_{10}\ ,\ \ \ \ \ 
N_\rho^+ - N_\rho ^- = \ell_{11}
\mbox{.}
\eqno{(5.4)}
$$
Here

$$
N^+_\sigma = \sum _{\ell=-\infty }^\infty \sum _{k=1}^\infty 
\frac{k}{\omega_{k\ell} }
\alpha^i_{(-k,-m)} \alpha^i_{(k,m)}
\mbox{,}
\eqno{(5.5)}
$$
$$
N^-_\sigma = \sum _{\ell=-\infty }^\infty \sum _{k=1}^\infty \frac{k}
{\omega_{k\ell} }
\widetilde \alpha^i_{(-k,-\ell)} \widetilde \alpha^i_{(k,\ell)}
\mbox{,}
\eqno{(5.6)}
$$
$$
N^+_\rho=\sum _{\ell=1}^\infty \sum _{k=0}^\infty \frac{\ell}{\omega_{k\ell} }
\big[ \alpha^i_{(-k,-\ell)} \alpha^i_{(k,\ell)} + 
\widetilde \alpha^i_{(-k,\ell)} \widetilde \alpha^i_{(k,-\ell)} \big]
\mbox{,}
\eqno{(5.7)}
$$
$$
N^-_\rho=\sum _{\ell=1}^\infty \sum _{k=0}^\infty \frac{\ell}{\omega_{k\ell} }
\big[ \alpha^i_ {(-k,\ell)}\alpha^i_{(k,-\ell)} + 
\widetilde \alpha^i_{(-k,-\ell)} \widetilde \alpha^i_{(k,\ell)} \big]
\mbox{.}
\eqno{(5.8)}
$$
Let us now define the quantum operator $H$ as 
$$
{\widehat H}= \sum _{\bf n} \big(: \alpha^i_{(-k,-\ell)} 
\alpha^i_{ (k,\ell)} : + :\widetilde \alpha^i_{(-k,-\ell)} 
\widetilde \alpha^i_{ (k,\ell) }:\big)
\mbox{,}
\eqno{(5.9)}
$$
where the annihilation operators $\alpha _{ {(k,\ell)} }^i, \ 
\widetilde \alpha _{ {(k,\ell)}}^i$ are determined for $k>0$ and 
$\ell\in {\mathbb Z}$, and $k=0$, $\ell>0$. In Eq. (5.9) the normal ordering
means taking the annihilation operators to the right. One can find the
relation (see Ref. \cite{russo})

$$
H=  {\widehat H}+ 2(D-3)E,\,\,\,\,\,\,\,\,\,\,
E= \frac{1}{2} \sum_{k,\ell} \omega_{k\ell}
\mbox{.}
\eqno{(5.10)}
$$
The constant energy shift $2(D-3)E$\,\,\,($E$ is the Casimir energy) represents 
the purely bosonic contribution to the vacuum energy of the (2+1) dimensional 
field theory. In the case of supersymmetry preserving boundary conditions for 
fermions the contributions to the vacuum energy coming from bosonic and fermionic 
fields cancel out \cite{berg1,duff}. This result also holds in  a supersymmetryic
theory when non-linear terms are included.

\subsection{Finite temperature quantum states in M-theory}

Here we consider membrane excitation states with non-trivial
winding numbers around the target space torus. It can be shown that the 
spectrum of the light-cone membrane Hamiltonian is discrete 
\cite{berg1,duff, russo}. Let the Euclidean time coordinate 
$X^0$ plays the role of $X^{10}$. Then fermions will obey antiperiodic boundary 
conditions around $X^0$ but periodic boundary conditions around $X^{11}$.
In the sector $m_0=\pm 1$ fermions are antiperiodic under the replacement 
$\sigma\rightarrow\sigma+2\pi$ (while periodic under $
\rho\rightarrow\rho+2\pi$). The Hamiltonian operator becomes

$$
{\mathfrak H}= \frac{\ell^2_{0}}{R_{0}^2} + \frac{\ell^2_{11}}{R_{11}^2} + 
\frac{R_{0}^2}{\alpha  ^{\prime 2}}\  + \frac{1}{\alpha' } ({\widehat H} + 
2(D-3)E)
\mbox{,}
\eqno{(5.11)} 
$$
where 
$$
{\widehat H}=  \sum _{\bf n} \big[ :\alpha^i_{-{\bf n}} \alpha^i_{{\bf n}} 
:+ :\widetilde \alpha^i_{-{\bf n}} \widetilde \alpha^i_{{\bf n}}: +
\omega_{k+ \frac{1}{2} ,\ell} \big( :S^a_{-{\bf n}} S^a_{{\bf n}} : 
+ :\widetilde S^a_{-{\bf n}} \widetilde S^a_{{\bf n}}:\big)\big]
\mbox{,}
\eqno{(5.12)}
$$
and 
$$
E= E_{\rm B}+ E_{\rm F}= \frac{1}{2} 
\sum_{k,m} \big( \omega_{km}- \omega_{k+\frac{1}{2}, m} \big),\,\,\,\,\,
\omega_{k\ell}=\bigg({k^2} +\frac{\ell^2}{{\rm g}_{eff} ^2} \bigg)^{1/2}
\mbox{.}
\eqno{(5.13)}
$$

In the case of supersymmetric membrane Eq. (2.18) gives

$$
b_{sp}(2)= \frac{3}{2}\left[28\zeta(3)A(2)\right]^{1/3}
\mbox{.}
\eqno{(5.14)}
$$

Eq. (2.16) shows that for linear Regge-like trajectories the termal
partition function always diverges ($\int dM\exp(M^{3/4}-\beta M)$ is 
divergent). This IR divergence in the free energy might be regularized
by some effects of brane theory, for example, like imposing U-duality
or choosing non-linear behavior of Regge trajectory (let say $M^{(1+p)/p}$ 
or something similar). The U-duality
properties of the membrane, considered in this paper, has been discussed
in Ref. \cite{bytsenko93}. Even in the divergent case the factor $b_{sp}(p)$, 
associated with the regularized partition function, 
gives a correct value of the brane critical temperature.   
The statistical mechanical density of states (degeneracies) is given in
Eqs. (2.13), (2.16) - (2.18). In the supersymmetric case the prefactors
$C(2)=C_{-}(2)C_{+}(2)$\, ($C_{-}(2)$, $C_{+}(2)$ given by Eq. (2.14))
represent general quantum field theoretical correction to the state density.

Following the lines of paper \cite{russo} and taking into account that 
${\bf a}={\rm diag}(1, {\rm g}_{\rm eff}^{-2})$ and
$A(2)=2\pi{\rm g}_{\rm eff}$,\, we havethe critical value 
$\beta_{cr}\simeq b_{sp}(2) = \,\,
(3/2)\left[56\pi \zeta(3){\rm g}_{\rm eff}\right]^{1/3}$.
Finally, restoring the dependence on parameter $\alpha'$ we have

$$
T_{\rm cr}\simeq \frac{2}{3\sqrt{\alpha '}}
\left[56\pi \zeta(3){\rm g}_{\rm eff}\right]^{-1/3}
\mbox{.}
\eqno{(5.15)}
$$
We would like remark that in this approach the folowing condition for coupling 
constants holds:  ${\rm g}_{\rm eff}^2=4\pi^2T_{cr}^2{\rm g}_A^2=
8\pi^3\alpha'T_{cr}^2R_{11}^3(\ell_P)^{-1}$. Thus the critical 
temperature depends on the cutt-off parameter which is proportional to the Planck 
length $\ell_P$ \cite{byts3,russo}.

A more interesting possibility allows for a finite temperature be introduced into the 
quantized fundamental (super) $p-$ brane theory. We proof this statement in the next section.

\subsection{Field thermodynamics presented in M-theory}

In fact, the series (4.14), (4.15) are divergent, nevertheless one can construct an analytic continuation of 
these expressions. Let us define for $|z|<\infty$ two series

$$
{\mathfrak W}_{\pm}(z)=\sum_{k=0}^{\infty}\frac{\sqrt{\pi}}{\Gamma(k+1)
\Gamma\left(pk+\frac{p+2}{2}\right)}\nu_{\pm}(k;p)\left(\frac{z}{2}\right)^
{p(2k+1)+1}
\mbox{,}
\eqno{(5.16)}
$$
where the factors $\nu_{\pm}(k;p)$ have the form

$$
\nu_{-}(k;p)=(-1)^{pk+1}
\mbox{,}
\eqno{(5.17)}
$$
$$
\nu_{+}(k;p)=\nu_{-}(k;p)\left[1-2^{-p(2k+1)-1}\right]
\mbox{.}
\eqno{(5.18)}
$$
For finite variable $z$ these series converge and the convergence improves  
rapidly with the increasing of the integer number $p$. Let 
$z=j\cdot2\pi x$, then for the series

$$
\sum_{j=1}^{\infty}{\mathfrak W}_{\pm}(j\cdot2\pi x)=\sum_{j=1}^{\infty}
\sum_{k=0}^{\infty}\frac{\sqrt{\pi}\nu_{\pm}(k;p)}
{\Gamma(k+1)\Gamma\left(pk+\frac{p+2}{2}
\right)}(j\pi x)^{p(2k+1)+1}
\mbox{.}
\eqno{(5.19)}
$$
Now, if we commute the (now divergent) sum $\Sigma_j$ with $\Sigma_k$ extra terms of 
the type $x^{-1}W_{\pm}(p)$ wil be generated on the right hand side of equation (5.19). 
Thus the result is

$$
\sum_{j=1}^{\infty}{\mathfrak W}_{\pm}(j2\pi x) + x^{-1}W_{\pm}(p)
$$
$$
=\sum_{k=0}^{\infty}\frac{\pi}{\Gamma(k+1)\Gamma\left(
pk+\frac{p+2}{2}\right)}\nu_{\pm}(k;p)\zeta_{R}[-p(2k+1)]
x^{1+p(2k+1)}
$$
$$
=\sin\left(\frac{\pi p}{2}\right)\sum_{k=1}^{\infty}
\frac{\Gamma\left(pk+\frac{1-p}{2}\right)}{\Gamma(k)}\zeta_{\pm}(2pk+1-p)
x^{1+p(2k-1)}
\mbox{,}
\eqno{(5.20)}
$$
where $W_{\pm}(p)$ is an integer function of $p$ (see for example Ref. \cite{actor}). 
In the second equality the functional equation for $\zeta_{R}(s)$ has been  used. 

The new form of the free energy is: 

$$
F_{p}(\beta)\simeq \frac{Q(D,p)}{\sin\left(\frac{\pi p}{2}\right)}      
\sum_{k=0}^{\infty}\frac{(-1)^{pk}\pi}
{\Gamma(k+1)\Gamma\left(pk+\frac{p+2}{2}\right)}
$$
$$
\times
\zeta_{R}[-p(2k+1)]
\left(\frac{\beta}{y(p)}\right)^{-1-p(2k+1)}
\mbox{,}
\eqno{(5.21)}
$$

$$
F_{sp}(\beta)\simeq \frac{Q(D,p)}{\sin\left(\frac{\pi p}{2}\right)}      
\sum_{k=0}^{\infty}\frac{(-1)^{pk}\left[2-2^{-1-p(2k+1)}\right]\pi}
{\Gamma(k+1)\Gamma\left(pk+\frac{p+2}{2}\right)}
$$
$$
\times \zeta_{R}[-p(2k+1)]\left(\frac{\beta}{y(p)}\right)^
{-1-p(2k+1)}
\mbox{.}
\eqno{(5.22)}
$$ 
The divergent series in Eqs. (4.14) and (4.15) for the $p$-branes free 
energy, when reexpressed on the left hand side of Eq. (5.20), remain 
well-defined for finite temperature and have a smooth 
$\beta\rightarrow\infty\,\,(T\rightarrow 0)$ limit.
The statistical internal energy and the entropy of finite temperature field 
theories (5.21), (5.22) can be easy calculated using Eq. (2.5).

\section{Conclusions}

It has been demonstrated recently that the BPS part of spectrum of type IIB 
string on a circle does match with the BPS part of supermembrane spectrum of states. 
In this paper we had dealt with the same discrete supermembrane spectrum as it has 
been used in the membrane-string correspondence. We calculated
the critical temperature in the strong string coupling limit by considering
M-theory on ${\mathbb R}^9\otimes{\mathbb T}^2$ (one of the sides of the torus
is the Euclidean time direction, and fermions obey antiperiodic boundary 
conditions). Yet a finite temperature can be introduced in membrane thermodynamics; 
we have prooved this statement. It means physically that a membrane (if it can be 
quantized semi-classicaly) behaves like an ideal gas of quantum modes, which 
corresponds to a field theory at finite temperature (zero critical temperature).     

There are deep connections between strings and membranes; at least they 
should be considered as different limits of a more general M-theory. Indeed, 
string results may be obtained via membrane-string correspondence and vice 
versa. Therefore, even being not a fundamental theory of (super) $p-$ branes may 
provide new deep insights in the understanding of string theory and consistent 
formulation of M-theory.

\end{document}